\documentclass[%
 reprint,
 amsmath,amssymb,
 aps,
]{revtex4-1}

\usepackage{graphicx}
\usepackage{dcolumn}
\usepackage{bm}
\usepackage{color}
\RequirePackage[colorlinks=true
,urlcolor=blue
,anchorcolor=blue
,citecolor=blue
,filecolor=blue
,linkcolor=blue
,menucolor=blue
,linktocpage=true
,pdfproducer=medialab
,pdfa=true
]{hyperref}

\begin{document}


\title{Non-singular and ghost-free infinite derivative gravity with torsion }

\author{\'Alvaro de la Cruz-Dombriz$^1$, Francisco Jos\'e Maldonado Torralba$^{1,2}$, Anupam Mazumdar$^{2}$}
\affiliation{%
$^1$Cosmology and Gravity Group, Department of Mathematics and Applied Mathematics, University of Cape Town, Rondebosch 7701, Cape Town, South Africa.\\
$^2$Van Swinderen Institute, University of Groningen, 9747 AG Groningen, The Netherlands.
}%

\begin{abstract}
We present the most general quadratic curvature action with torsion including {\it infinite covariant derivatives} and study its implications around the Minkowski background via the Palatini approach. Provided the torsion is {\it solely} given by the background axial field, the metric and torsion are shown to decouple, and both of them can be made ghost and singularity free for a fermionic source.
\end{abstract}

\maketitle

The theory of general relativity (GR), which respects diffeomorphism invariance, can be modified to incorporate the gauge structure of the Poincar\'e
group, provided a torsion field is added~\cite{Kibble:1961ba}. This is known as Poincar\'e Gauge gravity ({\it c.f.}~\cite{Shapiro}). Both GR and Poincar\'e gravity suffer from the short distance behavior at a classical level in terms of blackhole and cosmological singularities, known as the ultraviolet (UV) problem. In this paper our aim will be to construct an action which recovers GR and Poincar\'e theory of gravity in the infrared (IR), while ameliorating the UV behavior of both metric and torsion fields. 
Infinite derivative gravity (IDG) can potentially ameliorate the classical UV behavior of a metric theory of gravity~\cite{Tomboulis:1997gg,Biswas:2005qr, Modesto, Biswas:2011ar}. Different approaches of IDG have been made in the context of {\it teleparallel} gravity \cite{Koivisto1} and symmetric {\it teleparallel} gravity \cite{Koivisto2}. However, constructing a Poincar\'e gravity possessing a better UV behavior at a classical level remains very challenging. To the best of our knowledge, the systematic study of infinite derivative extensions of the Poincar\'e gravity has never been done before. The aim of this paper will be to construct an action including metric and torsion fields up to quadratic in curvature with infinite covariant derivatives.
{\bf Most general quadratic action with torsion:} We start with the most general covariant action of gravity with no prior assumptions on the connection. In order to obtain the {\bf{quadratic}} action let us consider perturbations around the Minkowski metric
$g_{\mu\nu}=\eta_{\mu\nu}+h_{\mu\nu}$,
where $\mu,\nu=0,1,2,3$ and we work with $(-,+++)$ signature. We stick to the terms up to quadratic in curvature ${\cal O}(h^2)$. Regarding the connection, the only requirement is that the metricity condition, i.e., $\nabla_{\rho}g_{\mu\nu}=0$, is fulfilled. With that, one can find a relation between the Levi-Civita connection $\Gamma$, and the general one $\widetilde{\Gamma}$, namely
$\widetilde{\Gamma}^{\rho}_{\,\,\mu\nu}\doteq\Gamma^{\rho}_{\,\,\mu\nu}+K^{\rho}_{\,\,\mu\nu} $,
where $K$ is the so-called {\bf contorsion} tensor,
$K_{\,\,\mu\nu}^{\rho}\doteq T_{\,\,\mu\nu}^{\rho}+T_{\mu\,\,\nu}^{\,\,\rho}+T_{\nu\,\,\mu}^{\,\,\rho}$, which is defined in terms of the {\bf torsion} tensor: $T_{\,\,\nu\rho}^{\mu}=\widetilde{\Gamma}_{\,\,\left[\nu\rho\right]}^{\mu}$, where the symbol $[\cdot\cdot]$ means antisymmetrization of the indices. 
Note that if the effects of torsion in the action are to be considered, the deviation from the Levi-Civita connection must be 
${\cal O}(h)$, i.e., $K\sim {\cal O}(h)$, which is compatible with experiment~\cite{constraints}. The most general action for metric and torsion, quadratic in both,  which generalizes both the metric~\cite{Biswas:2011ar,Biswas:2016etb} and the teleparallel actions \cite{Koivisto1}, and generalization of second order curvature invariants~\cite{Christensen}, will be of the form:
\begin{widetext}
\begin{equation}\label{GEQ}\footnotesize
S=\int {\rm d}^{4}x\sqrt{-g}\left[\frac{\widetilde{R}}{2}+\widetilde{R}_{\mu_{1}\nu_{1}\rho_{1}\sigma_{1}}\mathcal{O}_{\mu_{2}\nu_{2}\rho_{2}\sigma_{2}}^{\mu_{1}\nu_{1}\rho_{1}\sigma_{1}}\widetilde{R}^{\mu_{2}\nu_{2}\rho_{2}\sigma_{2}}+\widetilde{R}_{\mu_{1}\nu_{1}\rho_{1}\sigma_{1}}\mathcal{O}_{\mu_{2}\nu_{2}\rho_{2}}^{\mu_{1}\nu_{1}\rho_{1}\sigma_{1}}K^{\mu_{2}\nu_{2}\rho_{2}}+K_{\mu_{1}\nu_{1}\rho_{1}}\mathcal{O}_{\mu_{2}\nu_{2}\rho_{2}}^{\mu_{1}\nu_{1}\rho_{1}}K^{\mu_{2}\nu_{2}\rho_{2}}\right],
\end{equation}
\end{widetext}
where $\mathcal{O}$ denote differential operators containing covariant derivatives and the Minkowski metric $\eta_{\mu\nu}$. Also, the tilde 
$\,\widetilde{\,}\,$ represents the quantities calculated with respect to the total connection $\widetilde{\Gamma}$. 
Indeed, (\ref{GEQ}) is the most general action satisfying the aforementioned requirements, since operators acting on the left can always be integrated by parts to provide operators acting on the right plus total derivatives. Action (\ref{GEQ}) is captured by $46$ functions containing infinite covariant derivatives. Such functions reduce to $19$ when imposing the Bianchi identities and the total derivatives are taken into account, see Appendix. Further note that the usual Poincar\'e gauge gravity (including Einstein-Cartan gravity) can be recovered from \eqref{GEQ} provided one takes the local limit.
The linearized version of the action (\ref{GEQ}) around the Minkowski background can be written as
\begin{equation}\footnotesize
\label{free}
S_q=-\int {\rm d}^{4}x\sqrt{-g}\left(\mathcal{L}_{M}+\mathcal{L}_{MT}+\mathcal{L}_{T}\right)=S_{M}+S_{MT}+S_{T},
\end{equation}
where
{\small
\begin{eqnarray}
\label{L_M}
\mathcal{L}_{M}&=&\frac{1}{2}h_{\mu\nu}\Box a\left(\Box\right)h^{\mu\nu}+h_{\mu}^{\,\,\alpha}b\left(\Box\right)\partial_{\alpha}\partial_{\sigma}h^{\sigma\mu}+hc\left(\Box\right)\partial_{\mu}\partial_{\nu}h^{\mu\nu}
\nonumber
\\
&+&\frac{1}{2}h\Box d\left(\Box\right)h+h^{\lambda\sigma}\frac{f\left(\Box\right)}{\Box}\partial_{\sigma}\partial_{\lambda}\partial_{\mu}\partial_{\nu}h^{\mu\nu},\\
\label{mixed}
\mathcal{L}_{MT}&=&h\Box u\left(\Box\right)\partial_{\rho}K_{\,\,\,\,\,\sigma}^{\rho\sigma}+h_{\mu\nu}v_{1}\left(\Box\right)\partial^{\mu}\partial^{\nu}\partial_{\rho}K_{\,\,\,\,\,\sigma}^{\rho\sigma}
\nonumber
\\
&+&h_{\mu\nu}v_{2}\left(\Box\right)\partial^{\nu}\partial_{\sigma}\partial_{\rho}K^{\mu\sigma\rho}
+h_{\mu\nu}\Box w\left(\Box\right)\partial_{\rho}K^{\rho\mu\nu}, \\
\mathcal{L}_{T}&=&K^{\mu\sigma\lambda}p_{1}\left(\Box\right)K_{\mu\sigma\lambda}+K^{\mu\sigma\lambda}p_{2}\left(\Box\right)K_{\mu\lambda\sigma}+K_{\mu\,\,\rho}^{\,\,\rho}p_{3}\left(\Box\right)K_{\,\,\,\,\,\sigma}^{\mu\sigma}
\nonumber
\\
&+&K_{\,\,\nu\rho}^{\mu}q_{1}\left(\Box\right)\partial_{\mu}\partial_{\sigma}K^{\sigma\nu\rho}+K_{\,\,\nu\rho}^{\mu}q_{2}\left(\Box\right)\partial_{\mu}\partial_{\sigma}K^{\sigma\rho\nu}
\nonumber
\\
&+&K_{\mu\,\,\,\,\,\nu}^{\,\,\rho}q_{3}\left(\Box\right)\partial_{\rho}\partial_{\sigma}K^{\mu\nu\sigma}+K_{\mu\,\,\,\,\,\nu}^{\,\,\rho}q_{4}\left(\Box\right)\partial_{\rho}\partial_{\sigma}K^{\mu\sigma\nu}
\nonumber
\\
&+&K_{\,\,\,\,\,\rho}^{\mu\rho}q_{5}\left(\Box\right)\partial_{\mu}\partial_{\nu}K_{\,\,\,\,\,\sigma}^{\nu\sigma}+K_{\,\,\,\lambda\sigma}^{\lambda}q_{6}\left(\Box\right)\partial_{\mu}\partial_{\alpha}K^{\sigma\mu\alpha}
\nonumber
\\
&+&K_{\mu}^{\,\,\nu\rho}s\left(\Box\right)\partial_{\nu}\partial_{\rho}\partial_{\alpha}\partial_{\sigma}K^{\mu\alpha\sigma},
\label{L_T}
\end{eqnarray}
}
where $a,b,c,d,f,u,v_{1,2},w,p_{1,2,3},q_{1,2,3,4,5}$ and $s$ are functions of covariant {\it infinite derivatives} of $\Box=g^{\mu\nu}\nabla_\mu\nabla_\nu$ (see Appendix for a detailed discussion). Note that $\Box$ is a dimensionful quantity, since strictly speaking there is a scale $\Box_s =\Box/M_s^2$, where $M_s$ is the new scale at which gravity is modified in four dimensions with $M_s < M_{\rm Planck}=1.2\times 10^{19}$~GeV.  In order not to clutter our formulae, we shall suppress writing $M_s$. Furthermore, note that $\mathcal{L}_{M}$ in (\ref{L_M}) has only metric terms and coincides with the non-torsion case  Lagrangian~\cite{Biswas:2011ar}, as expected. On the other hand, $\mathcal{L}_{MT}$ in (\ref{mixed}) represents the mixed terms between metric and torsion, and $\mathcal{L}_{T}$ in (\ref{L_T}) only contains torsion terms. Thus, (\ref{free}) is the new most generalized linearized action of gravity without making any assumption about the choice of connection. To the best of our knowledge this is the first such a generalization within Poincar\'e theory of gravity. Furthermore, when $\Box/M_s^2\rightarrow 0$, then the terms involving the operators $u,~w$ tend to 0, recovering the local action, therefore yielding a Poincar\'e gauge gravity in the IR. In general, these two terms in \eqref{mixed} break the Poincar\'e invariance.

{\bf Field Equations:} We apply the Palatini formalism~\cite{Ferraris} to obtain the field equations, finding differences with respect to the torsion-free case in both the Einstein and Cartan equations.  We vary action (\ref{free}) with respect to the metric, i.e., $\delta_g /\delta  g^{\mu\nu}$, to find the Einstein Equations, and with respect to contorsion $K_{\,\,\nu\rho}^{\mu}$ to yield the Cartan Equations, i.e., $\delta_K/\delta K_{\,\,\nu\rho}^{\mu}$. Equations of motion derived from $\mathcal{L}_{M}$ when varying with respect to metric tensor have already been calculated in~\cite{Biswas:2011ar,Biswas:2013cha}, which serves as a consistency check for our calculations. It is worth noting that this is the first time the Palatini approach has been used in the context of IDG, since
the existing literature has always assumed the Levi-Civita as the underlying connection and hence the use of the metric formalism.
In the following we will sketch the calculations leading us to the generalized field equations.

{\bf Einstein Equations}: Variations with respect to the metric in $\mathcal{L}_{M}$ as presented in (\ref{mixed}) yield
{\footnotesize
\begin{eqnarray}
\label{eins1}
\frac{\delta_{g}S_{M}}{\delta g^{\mu\nu}}&=&\Box a\left(\Box\right)h_{\mu\nu}+b\left(\Box\right)\partial_{\sigma}\partial_{\left(\nu\right.}h_{\left.\mu\right)}^{\,\,\,\sigma}
\nonumber
\\
&+&c\left(\Box\right)\left[\partial_{\mu}\partial_{\nu}h+\eta_{\mu\nu}\partial_{\rho}\partial_{\sigma}h^{\rho\sigma}\right]+\eta_{\mu\nu}\Box d\left(\Box\right)h
\nonumber
\\
&+&\frac{2\,f\left(\Box\right)}{\Box}\partial_{\mu}\partial_{\nu}\partial_{\rho}\partial_{\sigma}h^{\rho\sigma},
\end{eqnarray}
}
an expression which is compatible with the results in~\cite{Biswas:2011ar}. $f(\Box)$ can be proved to have a polynomial form in $\Box$, so there are no inverse, non-analytic  $1/\Box$ operators involved in \eqref{eins1}. From the explicit expression of the functions in \eqref{eins1}, the following relations can be obtained~\cite{Biswas:2011ar} 
{\footnotesize
\begin{equation}\label{const-0}
a(\Box)+b(\Box)=0,~c(\Box)+d(\Box)=0,~b(\Box)+c(\Box)+f(\Box)=0\,,
\end{equation}
}
which are a consequence of the Bianchi identities and the conservation of the energy-momentum tensor. 
For $\mathcal{L}_{MT}$ we have
{\footnotesize
\begin{eqnarray}
\label{eins2}
\frac{\delta_{g}S_{MT}}{\delta g^{\mu\nu}}&=&\eta_{\mu\nu}\Box u\left(\Box\right)\partial_{\rho}K_{\,\,\,\,\,\sigma}^{\rho\sigma}+v_{1}\left(\Box\right)\partial_{\mu}\partial_{\nu}\partial_{\rho}K_{\,\,\,\,\,\sigma}^{\rho\sigma}
\nonumber
\\
&+&v_{2}\left(\Box\right)\partial_{\sigma}\partial_{\rho}\partial_{\left(\nu\right.}K_{\left.\mu\right)}^{\,\,\,\sigma\rho}
+\Box w\left(\Box\right)\partial_{\rho}K_{\,\,\left(\mu\nu\right)}^{\rho}.
\end{eqnarray}
}
Again, from the explicit decomposition of the functions in \eqref{eins2}
(see Appendix), one obtains
{\small
\begin{equation}
u(\Box)+v_{1}(\Box)=0\,,~~~~~~v_{2}(\Box)-w(\Box)=0\,,
\end{equation}
}
where the second constraint above arises from the conservation of the energy-momentum tensor. Interestingly, when  
$u(\Box)=v_{1}(\Box)=v_{2}(\Box)=w(\Box)=0$,
the mixed term ${\cal L}_{MT}$ vanishes. Note that the contributions of the contorsion tensor in \eqref{eins2} are purely {\it symmetric}. In the next section, this property will allow us to obtain 
solutions able to ameliorate the classical UV behavior.

{\bf Cartan Equations}: Variations now with respect to the contorsion in (\ref{mixed}) and (\ref{L_T}) yield
{\footnotesize
\begin{eqnarray}
\label{cartan1}
\frac{\delta\mathcal{L}_{MT}}{\delta K_{\,\,\nu\rho}^{\mu}}&=&-\Box u\left(\Box\right)\partial^{\left[\nu\right.}\eta_{\left.\mu\right]}^{\rho}h-v_{1}\left(\Box\right)\partial^{\alpha}\partial^{\beta}\partial^{\left[\nu\right.}\eta_{\left.\mu\right]}^{\rho}h_{\alpha\beta}
\nonumber
\\
&+&v_{2}\left(\Box\right)\partial^{\beta}\partial^{\rho}\partial^{\left[\nu\right.}h_{\left.\mu\right]\beta}+\Box w\left(\Box\right)\partial_{\left[\mu\right.}h^{\left.\nu\right]\rho},\\
\label{cartan2}
\frac{\delta\mathcal{L}_{T}}{\delta K_{\,\,\nu\rho}^{\mu}}&=&2p_{1}\left(\Box\right)K_{\mu}^{\,\,\nu\rho}+2p_{2}\left(\Box\right)K_{\left[\mu\right.}^{\,\,\,\,\,\left.\rho\right]\nu}+2p_{3}\left(\Box\right)\eta^{\nu\left[\rho\right.}K_{\left.\mu\right]\,\,\,\,\,\sigma}^{\,\,\,\,\sigma}
\nonumber
\\
&-&2q_{1}\left(\Box\right)\partial_{\sigma}\partial_{\left[\mu\right.}K^{\left.\rho\right]\nu\sigma}+2q_{2}\left(\Box\right)\partial_{\sigma}\partial_{\left[\mu\right.}K^{\sigma\left|\rho\right]\nu}
\nonumber
\\
&+&q_{3}\left(\Box\right)\left(\partial^{\nu}\partial_{\sigma}K_{\left[\mu\right.}^{\,\,\,\left.\rho\right]\sigma}+\partial_{\sigma}\partial^{\left[\rho\right.}K_{\left.\mu\right]\,\,\,\,\,}^{\,\,\,\,\sigma\nu}\right)
\nonumber
\\
&+&2q_{4}\left(\Box\right)\partial^{\nu}\partial_{\sigma}K_{\mu}^{\,\,\,\sigma\rho}+2q_{5}\left(\Box\right)\eta^{\nu\left[\rho\right.}\partial_{\left.\mu\right]}\partial_{\lambda}K_{\,\,\,\,\,\sigma}^{\lambda\sigma}
\nonumber
\\
&+&q_{6}\left(\Box\right)\left(\partial_{\lambda}\partial_{\alpha}\eta_{\left[\mu\right.}^{\nu}K^{\left.\rho\right]\lambda\alpha}-\partial^{\nu}\partial^{\left[\rho\right.}K_{\left.\mu\right]\lambda}^{\,\,\,\,\,\,\lambda}\right)
\nonumber
\\
&+&2s\left(\Box\right)\partial^{\sigma}\partial^{\lambda}\partial^{\rho}\partial^{\left[\nu\right.}K_{\left.\mu\right]\sigma\lambda}\nonumber
\end{eqnarray}
}
Note that in the standard Poincar\'e gravity there are no mixed terms. However, in our case the presence of differential operators causes the emergence of mixed terms between the metric and the torsion. While applying the Palatini formalism there will be terms in the Cartan equations which are non-zero even if the torsion is set null. If we set the torsion to zero, the Cartan equations do not play any role in the dynamics of the system, and variations with respect to the contorsion cannot be performed. Hence, the sole remaining equations would be the usual Einstein equations in the IDG, which takes the same form as in~\cite{Biswas:2011ar}.      

{\bf Solutions:} In order to obtain solutions for the classes of theories provided by the action \eqref{free}, let us recall the torsion tensor property 
thanks to which such a tensor can be uniquely decomposed as~\cite{Shapiro}
\begin{equation}
\label{decompose}
T_{\mu\nu\rho}=\frac{1}{3}\left(T_{\nu}g_{\mu\rho}-T_{\mu}g_{\mu\nu}\right)-\frac{1}{6}\varepsilon_{\mu\nu\rho\sigma}S^{\sigma}+q_{\mu\nu\rho},
\end{equation}
where the components above are given by the {\bf Trace} vector: $T_{\mu}=T_{\,\,\mu\nu}^{\nu}$, the {\bf Axial} vector: $S^{\mu}=\varepsilon^{\rho\sigma\nu\mu}T_{\rho\sigma\nu}$, and a tensor: $q_{\,\,\nu\rho}^{\mu}$ such that $q_{\,\,\mu\nu}^{\nu}=0$ with $\varepsilon^{\rho\sigma\nu\mu}q_{\rho\sigma\nu}=0$, where $\varepsilon$ denotes the totally antisymmetric tensor in four dimensions. 
Making use of this decomposition, we find that the only components contributing in the Einstein equations turn out to be the trace and the tensor parts. Therefore, if we assume that the torsion field only possesses a non-vanishing axial component, the Einstein equations would reduce to~\cite{Biswas:2011ar}. This means that the metric solutions would be the same ones as in standard IDG. Nevertheless, Cartan equations still need to be solved for those metrics in order to obtain the torsion field solutions. One possible solution would then be:
\begin{equation}
u\left(\Box\right)=v_{1}\left(\Box\right)=v_{2}\left(\Box\right)=w\left(\Box\right)=0\,,
\label{sol1}
\end{equation}
which yields $\mathcal{L}_{MT}=0$, i.e., the metric and torsion fields are decoupled. As a consequence, in this scenario \eqref{sol1} the degrees of freedom can be studied for the metric and the Cartan theories separately.  Consequently the theory space reduces to that of the Poincar\'e gauge gravity.  Now let us study the conditions for which the Field Equations do not host extra dynamical degrees of freedom.
{\bf Ghost-free conditions for metric and torsion:} The equations of motion for the pure metric theory are given by (\ref{eins1}). Since there are infinite covariant derivatives are present, leads to the emergence of new dynamical degrees of freedom, including ghosts. It has been shown in \cite{Biswas:2011ar} that in order to ensure the metric part of the theory to be ghost-free with the same on/off-shell degrees of freedom as that of  the massless graviton in four-dimensions, we would require~\cite{Biswas:2011ar}
\begin{eqnarray}\label{erf} 
a(\Box)=c(\Box)={\rm e}^{\gamma(\Box)}\,, 
\end{eqnarray}
where $\gamma(\Box)$ is an {\it entire function}, which has no poles, suggesting that $a(\Box)$
does not introduce any new dynamical degrees of freedom~\cite{Biswas:2011ar}. We can also show that the form factor $
\widetilde{F}_3(\Box)$ in (\ref{lagrangian}) becomes redundant due to the fact that the Weyl part does not contribute at the background level around the  Minkowski spacetime.  The simplest choice would be to consider $\gamma(\Box)=\Box/M_s^2$. Also, the above expression (\ref{erf}) for $a(\Box)$ appears in the graviton propagator. The gauge independent part of the graviton propagator can be recast  in terms of the spin projection operators, i.e., spin-2, $P^{(2)}$, and spin-0, $P^{(0)}$~\cite{Biswas:2011ar}, 
{\small
\begin{equation}
\Pi(k^2)=\frac{1}{a(k^2)}\left[\frac{P^{(2)}}{k^2}-\frac{P^{(0)}}{2k^2}\right]=\frac{1}{{\rm e}^{\gamma(k^2)}}\,\Pi(k^2)^{(\rm GR)}\,,
\end{equation}
}
where spacetime indices have been suppressed. As a consequence, provided $a(k^2)$ is given by an exponential of an {\it entire function}, then it does not introduce any new pole, nor any new dynamical degree of freedom, and therefore the true dynamical degrees of freedom remains that of the massless GR.
In order to obtain the ghost-free conditions for the torsion axial vector,  the relevant Lagrangian $\mathcal{L}_{T}$ in \eqref{free} can be rewritten as
\begin{equation}
\label{LT_new}
\mathcal{L}_{T}=S_{\mu}\Box\Lambda\left(\Box\right)S^{\mu}-S_{\mu}\Sigma\left(\Box\right)\partial^{\mu}\partial_{\nu}S^{\nu},
\end{equation}
where $\Sigma\left(\Box\right)=q_{1}\left(\Box\right)-q_{2}\left(\Box\right)-q_{3}\left(\Box\right)+q_{4}\left(\Box\right)$, and $\Lambda\left(\Box\right)=3\left(p_{1}\left(\Box\right)+p_{2}\left(\Box\right)\right)+\Sigma\left(\Box\right)$.
Hence the Cartan equations become
\begin{equation}
\label{cartan3}
\Box\Lambda\left(\Box\right)S^{\mu}-\Sigma\left(\Box\right)\partial^{\mu}\partial_{\nu}S^{\nu}=0.
\end{equation}
Consequently, now the torsion propagator $\mathcal{O}$ can be recast in terms of its corresponding degrees of freedom, namely the spin-0 and spin-1 modes, as follows
\begin{equation}
\mathcal{O}(k^2)=\frac{P^{(0)}}{\Lambda\left(-k^{2}\right)-\Sigma\left(-k^{2}\right)}+\frac{P^{(1)}}{-k^{2}\Lambda\left(-k^{2}\right)}.
\end{equation}
In order to have a ghost-free axial vector field, we need to impose that the scalar mode does not propagate, so only the spin-1 component propagates, and both $\Lambda$ and $\Sigma$ in \eqref{LT_new} must be of the form of an exponential of an entire function in order not to introduce any new degrees of freedom. Therefore, we have
\begin{equation}
\label{beta}
\Lambda\left(-k^{2}\right)=\Sigma\left(-k^{2}\right)={\rm e}^{\beta(k^2)}\,,
\end{equation}
where $\beta $ is an entire function, which introduces neither new poles nor new degrees of freedom.

\begin{figure}
\begin{center}
\includegraphics[width=0.6\linewidth]{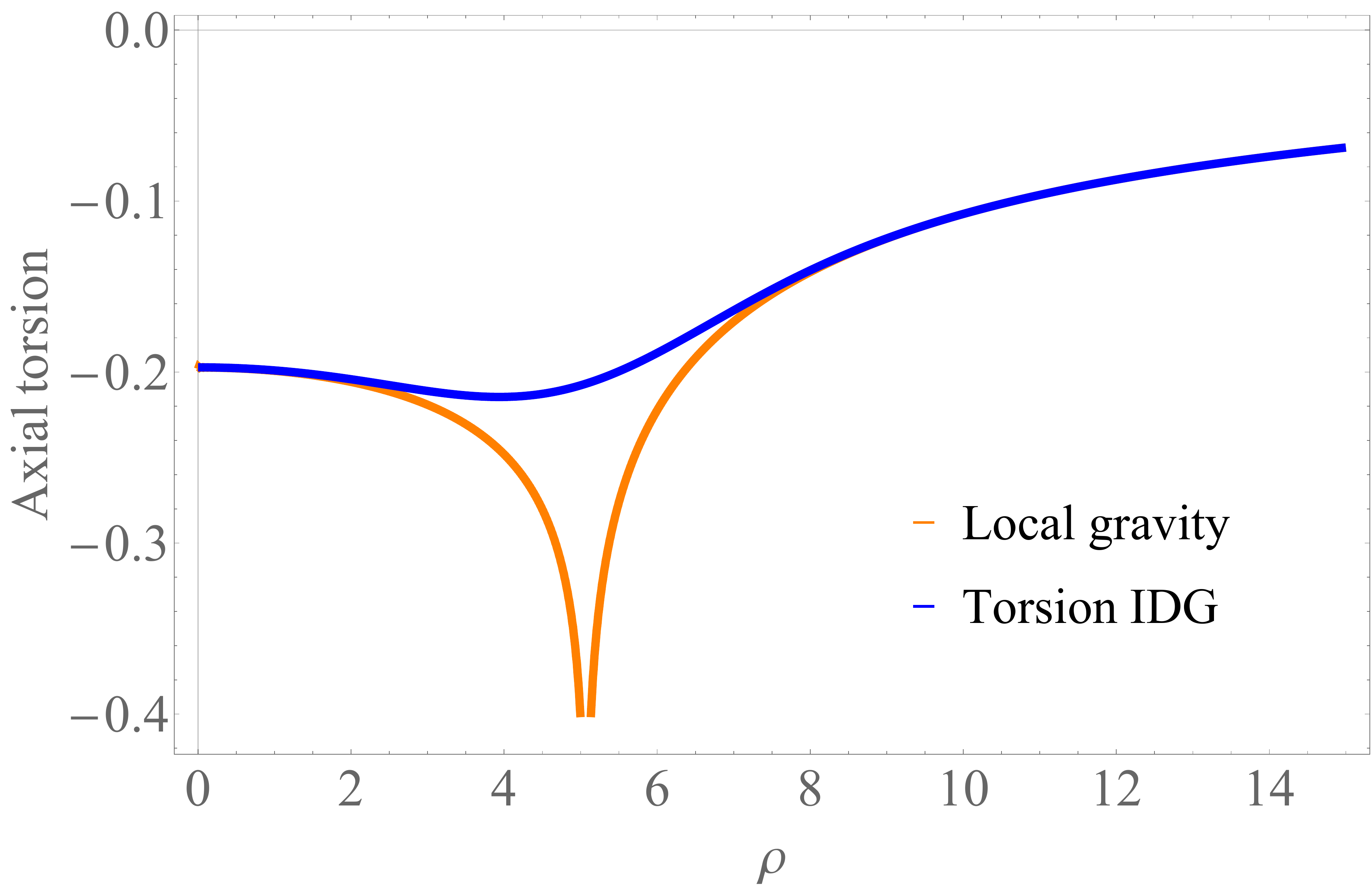}
\par\end{center}
\caption{Results of the numerical computation of \eqref{integral} for the case of local theories of gravity (limit when $M_{s}\rightarrow \infty$ or $\Box/M_s\rightarrow 0$) and IDG theories with torsion. We have chosen $A^{\mu}=4,\,\mu=\left\{1,...,4\right\}$, $R=5.06$, and $M_{s}=1$. }
\label{figura}
\end{figure}


\paragraph*{\bf Avoidance of point-source singularity:} The simplest choice of both entire functions $\beta$ and $\gamma$ would be $\beta(k^2) \sim \gamma(k^2)=k^2/M_s^2$. Other choices for these entire functions can also be made~\cite{Edholm:2016hbt}, but neither the UV nor the IR part are overtly sensitive enough to such choices. Let us first deal with the metric theory of gravity, where it has already been shown that in presence of a massive static point source, i.e., the $00$ component of the energy-momentum tensor can be written as: $\tau_{00}=m \delta^{(3)}(r)$. Accordingly the Einstein equations (\ref{eins1}) along with (\ref{const-0}) yield a non-singular solution in isotropic coordinates, 
\begin{eqnarray}
{\rm ds}^2=-(1+2\Phi){\rm d}t^2+(1-2\Psi){\rm d}\vec r^{\,2}\,, \label{iso}\\
\Phi(r)=\Psi(r)=-\frac{Gm}{r}\,{\rm Erf}(rM_s/2)\,,
\end{eqnarray}
where $G$ is the Newton's constant, and one can see that when $r\gg 2/M_s$, we recover the correct IR limit of a metric theory of gravity, i.e., the correct Newtonian limit of massless gravity, while in the UV counterpart, i.e., $r< 2/M_s$,  the metric potentials approach to be constant, $\Phi=\Psi= GmM_s/\sqrt{\pi}$. In fact, it has been shown that in UV limit the Weyl tensor vanishes linearly in $r$, and both the Ricci scalar and the Ricci tensor approach to constant values. Effectively, gravity becomes conformally flat in the UV limit~\cite{Buoninfante:2018xiw}. The time-dependent dynamical equations, such as those in the matter collapse case, have also been solved and shown to be non-singular~\cite{Frolov:2015bia}. Analogous non-singular solutions exist in three~\cite{Mazumdar:2018xjz} and higher dimensions~\cite{Boos:2018bxf}. The complete nonlinear equations of motion do not permit singular behaviour $r^{-\alpha}$, ($\alpha>0$)~\cite{Koshelev:2018hpt}, and it is possible that astrophysical objects can be made devoid of singularity as well as horizons~\cite{Koshelev:2017bxd,Buoninfante:2018xiw}.

Now, let us study the evolution of the axial torsion field $S^{\mu}$, which couples to fermionic sources only. Since fermions have an intrinsic spin, instead of having a Dirac-delta point source for a fermion, we would need to consider a singular source endowed with angular momentum. Indeed, for the sake of simplicity we can fix the angular momentum to be in $z$ direction, without any loss of generality. However, a mere Dirac-delta at $r=0$ will not be able to capture the spin of the fermion, instead we would need a rotating singular Dirac-delta ring. We will use again isotropic coordinates (\ref{iso}), in which the Cartan equations \eqref{cartan3} become
{\small
\begin{equation}
\label{singular}
\Box\,{\rm e}^{\beta(\Box)}S^{\mu}=A^{\mu}\delta\left(z\right)\delta\left(x^{2}+y^{2}-R^{2}\right),
\end{equation}
}
where $A^{\mu}$ holds for a source vector and $R$ for the constant Cartan radius of a singular rotating ring, where effectively the singularity is located. For illustrative purposes, we may assume 
$\beta(\Box)={\Box}/{M_{s}^{2}}$. In order to solve Eq. \eqref{singular}, we need to calculate the Fourier transform $\mathcal{F}$ of the source, as follows
{\small
\begin{equation}
\mathcal{F}\left[\delta\left(z\right)\delta\left(x^{2}+y^{2}-R^{2}\right)\right]=\pi {\rm J}_{0}\left(-R\sqrt{k_{x}^{2}+k_{y}^{2}}\right),
\end{equation} 
}
where ${\rm J}_{0}$ represents the Bessel function of first kind ($n=0$). Thus, the solution of Eq.\eqref{singular} can be expressed as
{\small
\begin{eqnarray}
\label{sol_S}
S^{\mu}=-\pi A^{\mu}&\int_{}^{}&\frac{{\rm d}^{3}k}{\left(2\pi\right)^{3}}\,{\rm e}^{-\frac{k^{2}}{M_{s}^{2}}}\,{\rm J}_{0}\left(-R\sqrt{k_{x}^{2}+k_{y}^{2}}\right)
\nonumber
\\
&&\times\, {\rm e}^{i\left(k_{x}x + k_{y}y + k_{z}z\right)},
\end{eqnarray}
}
where ${\rm d}^{3}k={\rm d}k_{x}{\rm d}k_{y}{\rm d}k_{z}$ and $k^{2}=k_{x}^{2}+k_{y}^{2}+k_{z}^{2}$. In order to see how the axial vector behaves at the singularity $r=R$, we can restrict the study of the integral in (\ref{sol_S}) to the $z=0$ plane, assuming that the ring rotation axis lies along the $z$ direction. By using cylindrical coordinates, $k_{x}=r\cos\left(\varphi\right)$, $k_{y}=r\sin\left(\varphi\right)$, $k_{z}=k_{z}$, we obtain
{\small
\begin{equation}
\label{integral}
S^{\mu}\left(\rho\right)=-\frac{1}{4}A^{\mu}\int_{0}^{\infty}{\rm d}r{\rm J}_{0}\left(-Rr\right){\rm J}_{0}\left(-r\rho\right){\rm Erfc}\left(r/{M_{s}}\right)\,,
\end{equation}
}
where ${\rm Erfc}(z)=1-{\rm Erf}(z)$ is the complementary error function. Since finding the analytically closed form is not possible,  the integral in (\ref{integral}) can be solved numerically for physically relevant values, as can be seen in Fig. \ref{figura}.  In the case of stable local Poincar\'e Gauge theories of gravity, in the limit $M_{s}\rightarrow \infty$, the singularity at $r=R$ is unavoidable, see \cite{delaCruz-Dombriz:2018vzn}. Within the infinite derivative theory of Poincar\'e gravity, which has no ghosts, the ring singularity can be smeared out. In both cases in the figure, the axial torsion presents a $1/r$ behavior in the IR limit. Therefore, for IDG theories we conclude that the axial torsion is regular everywhere in presence of a Dirac-delta fermionic source with spin. This result is similar to the Kerr-like singularity which is cured  in the infinite derivative metric theory of gravity~\cite{Buoninfante:2018xif}.

\paragraph*{\bf Conclusions:} In this paper, we have presented the most general action for infinite derivative gravity with torsion. We have applied the Palatini formalism for the first time in this context to obtain the linearized field equations around the Minkowski background. Under the assumption that the torsion component is solely given by an axial field, we have shown that the coupling between the metric and the torsion vanishes, therefore leading to two separate infinite derivative theories of gravity along with an axial torsion field. For both sectors, we were able to show that the ghost-free conditions are able to smear out point and ring-like singularities. The solution where the axial torsion couples to a fermionic source with a spin is absolutely novel. Consequently, our results may have seminal consequences for building quantum theory of gravity with spins and understanding the UV aspects of Poincar\'e gauge theories of gravity.

{\bf Acknowledgements}
 AM's research is funded by the Netherlands Organization for Scientific Research (NWO) grant number 680-91-119. AdlCD and FJMT acknowledge financial support from UCT Launching Grants Programme and NRF Grants No. 99077 2016-2018, Ref. No. CSUR150628121624, 110966 Ref. No. BS170509230233, and the NRF IPRR, Ref. No. IFR170131220846.

\paragraph*{\bf Appendix:}The full non-linear quadratic Lagrangian in both curvature and torsion (around Minkowski) is given by

\begin{widetext}
{\scriptsize
\begin{eqnarray}
\label{lagrangian}
\mathcal{L}_{q}&=&\widetilde{R}\widetilde{F}_{1}\left(\Box\right)\widetilde{R}+\widetilde{R}\widetilde{F}_{2}\left(\Box\right)\partial_{\mu}\partial_{\nu}\widetilde{R}^{\mu\nu}+\widetilde{R}_{\mu\nu}\widetilde{F}_{3}\left(\Box\right)\widetilde{R}^{\left(\mu\nu\right)}+\widetilde{R}_{\mu\nu}\widetilde{F}_{4}\left(\Box\right)\widetilde{R}^{\left[\mu\nu\right]}+\widetilde{R}_{\left(\mu\right.}^{\,\,\,\left.\nu\right)}\widetilde{F}_{5}\left(\Box\right)\partial_{\nu}\partial_{\lambda}\widetilde{R}^{\mu\lambda}+\widetilde{R}_{\left[\mu\right.}^{\,\,\,\left.\nu\right]}\widetilde{F}_{6}\left(\Box\right)\partial_{\nu}\partial_{\lambda}\widetilde{R}^{\mu\lambda}
\nonumber
\\
&+&\widetilde{R}_{\mu}^{\,\,\,\nu}\widetilde{F}_{7}\left(\Box\right)\partial_{\nu}\partial_{\lambda}\widetilde{R}^{\left(\mu\lambda\right)}+\widetilde{R}_{\mu}^{\,\,\,\nu}\widetilde{F}_{8}\left(\Box\right)\partial_{\nu}\partial_{\lambda}\widetilde{R}^{\left[\mu\lambda\right]}+\widetilde{R}^{\lambda\sigma}\widetilde{F}_{9}\left(\Box\right)\partial_{\mu}\partial_{\sigma}\partial_{\nu}\partial_{\lambda}\widetilde{R}^{\mu\nu}+\widetilde{R}_{\left(\mu\lambda\right)}\widetilde{F}_{10}\left(\Box\right)\partial_{\nu}\partial_{\sigma}\widetilde{R}^{\mu\nu\lambda\sigma}+\widetilde{R}_{\left[\mu\lambda\right]}\widetilde{F}_{11}\left(\Box\right)\partial_{\nu}\partial_{\sigma}\widetilde{R}^{\mu\nu\lambda\sigma}
\nonumber
\\
&+&\widetilde{R}_{\mu\lambda}\widetilde{F}_{12}\left(\Box\right)\partial_{\nu}\partial_{\sigma}\widetilde{R}^{\left(\mu\nu\right|\left.\lambda\sigma\right)}+\widetilde{R}_{\mu\lambda}\widetilde{F}_{13}\left(\Box\right)\partial_{\nu}\partial_{\sigma}\widetilde{R}^{\left[\mu\nu\right|\left.\lambda\sigma\right]}+\widetilde{R}_{\mu\nu\lambda\sigma}\widetilde{F}_{14}\left(\Box\right)\widetilde{R}^{\left(\mu\nu\right|\left.\lambda\sigma\right)}+\widetilde{R}_{\mu\nu\lambda\sigma}\widetilde{F}_{15}\left(\Box\right)\widetilde{R}^{\left[\mu\nu\right|\left.\lambda\sigma\right]}+\widetilde{R}_{\left(\rho\mu\right|\left.\nu\lambda\right)}\widetilde{F}_{16}\left(\Box\right)\partial^{\rho}\partial_{\sigma}\widetilde{R}^{\mu\nu\lambda\sigma}
\nonumber
\\
&+&\widetilde{R}_{\left[\rho\mu\right|\left.\nu\lambda\right]}\widetilde{F}_{17}\left(\Box\right)\partial^{\rho}\partial_{\sigma}\widetilde{R}^{\mu\nu\lambda\sigma}+\widetilde{R}_{\rho\mu\nu\lambda}\widetilde{F}_{18}\left(\Box\right)\partial^{\rho}\partial_{\sigma}\widetilde{R}^{\left(\mu\nu\right|\left.\lambda\sigma\right)}+\widetilde{R}_{\rho\mu\nu\lambda}\widetilde{F}_{19}\left(\Box\right)\partial^{\rho}\partial_{\sigma}\widetilde{R}^{\left[\mu\nu\right|\left.\lambda\sigma\right]}+\widetilde{R}_{\left(\mu\nu\right|\left.\rho\sigma\right)}\widetilde{F}_{20}\left(\Box\right)\partial^{\nu}\partial^{\sigma}\partial_{\alpha}\partial_{\beta}\widetilde{R}^{\mu\alpha\rho\beta}
\nonumber
\\
&+&\widetilde{R}_{\left[\mu\nu\right|\left.\rho\sigma\right]}\widetilde{F}_{21}\left(\Box\right)\partial^{\nu}\partial^{\sigma}\partial_{\alpha}\partial_{\beta}\widetilde{R}^{\mu\alpha\rho\beta}+\widetilde{R}_{\mu\nu\rho\sigma}\widetilde{F}_{22}\left(\Box\right)\partial^{\nu}\partial^{\sigma}\partial_{\alpha}\partial_{\beta}\widetilde{R}^{\left(\mu\alpha\right|\left.\rho\beta\right)}+\widetilde{R}_{\mu\nu\rho\sigma}\widetilde{F}_{23}\left(\Box\right)\partial^{\nu}\partial^{\sigma}\partial_{\alpha}\partial_{\beta}\widetilde{R}^{\left[\mu\alpha\right|\left.\rho\beta\right]}+K_{\mu\nu\rho}\widetilde{F}_{24}\left(\Box\right)K^{\mu\nu\rho}
\nonumber
\\
&+&K_{\mu\nu\rho}\widetilde{F}_{25}\left(\Box\right)K^{\mu\rho\nu}+K_{\mu\,\,\rho}^{\,\,\rho}\widetilde{F}_{26}\left(\Box\right)K_{\,\,\,\,\,\sigma}^{\mu\sigma}+K_{\,\,\nu\rho}^{\mu}\widetilde{F}_{27}\left(\Box\right)\partial_{\mu}\partial_{\sigma}K^{\sigma\nu\rho}+K_{\,\,\nu\rho}^{\mu}\widetilde{F}_{28}\left(\Box\right)\partial_{\mu}\partial_{\sigma}K^{\sigma\rho\nu}+K_{\mu\,\,\,\,\,\nu}^{\,\,\rho}\widetilde{F}_{29}\left(\Box\right)\partial_{\rho}\partial_{\sigma}K^{\mu\nu\sigma}
\nonumber
\\
&+&K_{\mu\,\,\,\,\,\nu}^{\,\,\rho}\widetilde{F}_{30}\left(\Box\right)\partial_{\rho}\partial_{\sigma}K^{\mu\sigma\nu}+K_{\,\,\,\,\,\rho}^{\mu\rho}\widetilde{F}_{31}\left(\Box\right)\partial_{\mu}\partial_{\nu}K_{\,\,\,\,\,\sigma}^{\nu\sigma}+K_{\mu}^{\,\,\nu\rho}\widetilde{F}_{32}\left(\Box\right)\partial_{\nu}\partial_{\rho}\partial_{\alpha}\partial_{\sigma}K^{\mu\alpha\sigma}+K_{\,\,\,\lambda\sigma}^{\lambda}\widetilde{F}_{33}\left(\Box\right)\partial_{\rho}\partial_{\nu}K^{\nu\rho\sigma}+\widetilde{R}_{\,\,\nu\rho\sigma}^{\mu}\widetilde{F}_{34}\left(\Box\right)\partial_{\mu}K^{\nu\rho\sigma}
\nonumber
\\
&+&\widetilde{R}_{\mu\nu\,\,\sigma}^{\,\,\,\,\,\,\rho}\widetilde{F}_{35}\left(\Box\right)\partial_{\rho}K^{\mu\nu\sigma}+\widetilde{R}_{\left(\rho\sigma\right)}\widetilde{F}_{36}\left(\Box\right)\partial_{\nu}K^{\nu\rho\sigma}+\widetilde{R}_{\left[\rho\sigma\right]}\widetilde{F}_{37}\left(\Box\right)\partial_{\nu}K^{\nu\rho\sigma}+\widetilde{R}_{\rho\sigma}\widetilde{F}_{38}\left(\Box\right)\partial_{\nu}K^{\rho\nu\sigma}+\widetilde{R}_{\left(\rho\sigma\right)}\widetilde{F}_{39}\left(\Box\right)\partial^{\sigma}K_{\,\,\,\,\,\mu}^{\rho\mu}+\widetilde{R}_{\left[\rho\sigma\right]}\widetilde{F}_{40}\left(\Box\right)\partial^{\sigma}K_{\,\,\,\,\,\mu}^{\rho\mu}
\nonumber
\\
&+&\widetilde{R}\widetilde{F}_{41}\left(\Box\right)\partial_{\rho}K_{\,\,\,\,\,\mu}^{\rho\mu}+\widetilde{R}_{\,\,\alpha\,\,\sigma}^{\mu\,\,\rho}\widetilde{F}_{42}\left(\Box\right)\partial_{\mu}\partial_{\rho}\partial_{\nu}K^{\nu\left(\alpha\sigma\right)}+\widetilde{R}_{\,\,\alpha\,\,\sigma}^{\mu\,\,\rho}\widetilde{F}_{43}\left(\Box\right)\partial_{\mu}\partial_{\rho}\partial_{\nu}K^{\nu\left[\alpha\sigma\right]}+\widetilde{R}_{\,\,\alpha\,\,\sigma}^{\mu\,\,\rho}\widetilde{F}_{44}\left(\Box\right)\partial_{\mu}\partial_{\rho}\partial_{\nu}K^{\alpha\nu\sigma}+\widetilde{R}_{\,\,\left.\sigma\right)}^{\left(\mu\right.}\widetilde{F}_{45}\left(\Box\right)\partial_{\mu}\partial_{\nu}\partial_{\alpha}K^{\sigma\nu\alpha}
\nonumber
\\
&+&\widetilde{R}_{\,\,\left.\sigma\right]}^{\left[\mu\right.}\widetilde{F}_{46}\left(\Box\right)\partial_{\mu}\partial_{\nu}\partial_{\alpha}K^{\sigma\nu\alpha},
\end{eqnarray}
}
where {\footnotesize $\widetilde{R}_{\left(\alpha\beta\right|\left.\gamma\rho\right)}=\frac{1}{2}\left(\widetilde{R}_{\alpha\beta\gamma\rho}+\widetilde{R}_{\gamma\rho\alpha\beta}\right)$} and {\footnotesize $\widetilde{R}_{\left[\alpha\beta\right|\left.\gamma\rho\right]}=\frac{1}{2}\left(\widetilde{R}_{\alpha\beta\gamma\rho}-\widetilde{R}_{\gamma\rho\alpha\beta}\right)$}.
In order to be consistent with the action in~\cite{Biswas:2011ar} when the torsion is zero, we need to satisfy the following relations
{\scriptsize
\begin{equation}
\widetilde{F}_{5}\left(\Box\right)+\widetilde{F}_{7}\left(\Box\right)=F_{4}\left(\Box\right),~~~~\,\widetilde{F}_{10}\left(\Box\right)+\widetilde{F}_{12}\left(\Box\right)=F_{7}\left(\Box\right),~~~~\,\widetilde{F}_{16}\left(\Box\right)+\widetilde{F}_{18}\left(\Box\right)=F_{11}\left(\Box\right),~~~~\,
\widetilde{F}_{20}\left(\Box\right)+\widetilde{F}_{22}\left(\Box\right)=F_{12}\left(\Box\right).
\end{equation}
}
After some computations we arrive at Eq.\eqref{free}, where the coefficients therein are given by

{\scriptsize
\begin{eqnarray}
u\left(\Box\right)&=&-4\tilde{F}_{1}\left(\Box\right)-\tilde{F}_{5}\left(\Box\right)\Box-\tilde{F}_{7}\left(\Box\right)\Box-\tilde{F}_{9}\left(\Box\right)\Box^{2}+\frac{1}{2}\tilde{F}_{39}\left(\Box\right)+\tilde{F}_{41}\left(\Box\right),~\,\,\,v_{1}\left(\Box\right)=4\tilde{F}_{1}\left(\Box\right)+\tilde{F}_{5}\left(\Box\right)\Box+\tilde{F}_{7}\left(\Box\right)\Box+\tilde{F}_{9}\left(\Box\right)\Box^{2}-\frac{1}{2}\tilde{F}_{39}\left(\Box\right)-\tilde{F}_{41}\left(\Box\right),
\nonumber
\\ 
v_{2}\left(\Box\right)&=&-\frac{1}{2}\tilde{F}_{3}\left(\Box\right)-\tilde{F}_{10}\left(\Box\right)\Box-\tilde{F}_{12}\left(\Box\right)\Box+\tilde{F}_{9}\left(\Box\right)\Box^{2}-4\tilde{F}_{14}\left(\Box\right)-\tilde{F}_{16}\left(\Box\right)\Box-\tilde{F}_{18}\left(\Box\right)\Box-\tilde{F}_{20}\left(\Box\right)\Box^{2}-\tilde{F}_{22}\left(\Box\right)\Box^{2}+\frac{1}{2}\tilde{F}_{34}\left(\Box\right)+\frac{1}{2}\tilde{F}_{35}\left(\Box\right)
\nonumber
\\
&+&\frac{1}{2}\tilde{F}_{36}\left(\Box\right)+\frac{1}{2}\tilde{F}_{42}\left(\Box\right),~~~w\left(\Box\right)=-\frac{1}{2}\tilde{F}_{3}\left(\Box\right)-\tilde{F}_{10}\left(\Box\right)\Box-\tilde{F}_{12}\left(\Box\right)\Box+\tilde{F}_{9}\left(\Box\right)\Box^{2}-4\tilde{F}_{14}\left(\Box\right)-\tilde{F}_{16}\left(\Box\right)\Box-\tilde{F}_{18}\left(\Box\right)\Box-\tilde{F}_{20}\left(\Box\right)\Box^{2}
\nonumber
\\
&-&\tilde{F}_{22}\left(\Box\right)\Box^{2}+\frac{1}{2}\tilde{F}_{34}\left(\Box\right)+\frac{1}{2}\tilde{F}_{35}\left(\Box\right)+\frac{1}{2}\tilde{F}_{36}\left(\Box\right)+\frac{1}{2}\tilde{F}_{42}\left(\Box\right),~~~q_{1}\left(\Box\right)=\frac{1}{2}\tilde{F}_{3}\left(\Box\right)+\frac{1}{2}\tilde{F}_{4}\left(\Box\right)+\frac{1}{2}\tilde{F}_{10}\left(\Box\right)\Box+\frac{1}{2}\tilde{F}_{11}\left(\Box\right)\Box+\frac{1}{2}\tilde{F}_{12}\left(\Box\right)\Box
\nonumber
\\
&+&\frac{1}{2}\tilde{F}_{13}\left(\Box\right)\Box+\frac{1}{2}\tilde{F}_{16}\left(\Box\right)\Box+\frac{1}{2}\tilde{F}_{18}\left(\Box\right)\Box+\frac{1}{2}\tilde{F}_{19}\left(\Box\right)\Box+\frac{1}{2}\tilde{F}_{20}\left(\Box\right)\Box+\frac{1}{2}\tilde{F}_{21}\left(\Box\right)\Box+\frac{1}{2}\tilde{F}_{22}\left(\Box\right)\Box+\frac{1}{2}\tilde{F}_{23}\left(\Box\right)\Box+\tilde{F}_{27}\left(\Box\right)-\frac{1}{2}\tilde{F}_{36}\left(\Box\right)
\nonumber
\\
&-&\frac{1}{2}\tilde{F}_{37}\left(\Box\right)-\frac{1}{2}\tilde{F}_{42}\left(\Box\right)\Box-\frac{1}{2}\tilde{F}_{43}\left(\Box\right)\Box,~~~q_{2}\left(\Box\right)=\frac{1}{2}\tilde{F}_{3}\left(\Box\right)-\frac{1}{2}\tilde{F}_{4}\left(\Box\right)+\frac{1}{2}\tilde{F}_{10}\left(\Box\right)\Box-\frac{1}{2}\tilde{F}_{11}\left(\Box\right)\Box+\frac{1}{2}\tilde{F}_{12}\left(\Box\right)\Box-\frac{1}{2}\tilde{F}_{13}\left(\Box\right)\Box+2\tilde{F}_{14}\left(\Box\right)
\nonumber
\\
&-&2\tilde{F}_{15}\left(\Box\right)+\frac{1}{2}\tilde{F}_{16}\left(\Box\right)\Box+\frac{1}{2}\tilde{F}_{20}\left(\Box\right)\Box-\frac{1}{2}\tilde{F}_{21}\left(\Box\right)\Box+\frac{1}{2}\tilde{F}_{22}\left(\Box\right)\Box-\frac{1}{2}\tilde{F}_{23}\left(\Box\right)\Box+\tilde{F}_{28}\left(\Box\right)-\frac{1}{2}\tilde{F}_{36}\left(\Box\right)+\frac{1}{2}\tilde{F}_{37}\left(\Box\right)-\frac{1}{2}\tilde{F}_{42}\left(\Box\right)\Box+\frac{1}{2}\tilde{F}_{43}\left(\Box\right)\Box,
\nonumber
\\
q_{3}\left(\Box\right)&=&-\tilde{F}_{17}\left(\Box\right)\Box-\tilde{F}_{18}\left(\Box\right)\Box+\tilde{F}_{19}\left(\Box\right)\Box+\tilde{F}_{29}\left(\Box\right)+\tilde{F}_{34}\left(\Box\right)-\tilde{F}_{35}\left(\Box\right)-\tilde{F}_{38}\left(\Box\right)-\tilde{F}_{44}\left(\Box\right)\Box,~~~\,\,\,q_{4}\left(\Box\right)=-\tilde{F}_{14}\left(\Box\right)-\tilde{F}_{15}\left(\Box\right)+\tilde{F}_{30}\left(\Box\right),
\nonumber
\\
q_{5}\left(\Box\right)&=&4\tilde{F}_{1}\left(\Box\right)+2\tilde{F}_{2}\left(\Box\right)\Box+\frac{1}{2}\tilde{F}_{3}\left(\Box\right)-\frac{1}{2}\tilde{F}_{4}\left(\Box\right)+\tilde{F}_{5}\left(\Box\right)\Box+\tilde{F}_{7}\left(\Box\right)\Box+\tilde{F}_{9}\left(\Box\right)\Box^{2}+\tilde{F}_{31}\left(\Box\right)-\frac{1}{2}\tilde{F}_{39}\left(\Box\right)-\frac{1}{2}\tilde{F}_{40}\left(\Box\right)-2\tilde{F}_{41}\left(\Box\right),
\nonumber
\\
q_{6}\left(\Box\right)&=&\tilde{F}_{3}\left(\Box\right)+\tilde{F}_{4}\left(\Box\right)+\tilde{F}_{32}\left(\Box\right)+\frac{1}{2}\tilde{F}_{36}\left(\Box\right)+\frac{1}{2}\tilde{F}_{37}\left(\Box\right)-\tilde{F}_{38}\left(\Box\right)-\frac{1}{2}\tilde{F}_{39}\left(\Box\right)+\frac{1}{2}\tilde{F}_{40}\left(\Box\right)+\frac{1}{2}\tilde{F}_{45}\left(\Box\right)\Box+\frac{1}{2}\tilde{F}_{46}\left(\Box\right)\Box,
\nonumber
\\
p_{1}\left(\Box\right)&=&\tilde{F}_{14}\left(\Box\right)\Box+\tilde{F}_{15}\left(\Box\right)\Box+\tilde{F}_{24}\left(\Box\right),~~~p_{2}\left(\Box\right)=\frac{1}{2}\tilde{F}_{18}\left(\Box\right)\Box^{2}-\frac{1}{2}\tilde{F}_{19}\left(\Box\right)\Box^{2}+\tilde{F}_{25}\left(\Box\right)+\tilde{F}_{34}\left(\Box\right),~~~p_{3}\left(\Box\right)=\frac{1}{2}\tilde{F}_{3}\left(\Box\right)\Box+\frac{1}{2}\tilde{F}_{4}\left(\Box\right)\Box+\tilde{F}_{26}\left(\Box\right)
\nonumber
\\
&-&\frac{1}{2}\tilde{F}_{39}\left(\Box\right)\Box+\frac{1}{2}\tilde{F}_{40}\left(\Box\right)\Box,~~~s\left(\Box\right)=-\frac{1}{2}\tilde{F}_{10}\left(\Box\right)-\frac{1}{2}\tilde{F}_{11}\left(\Box\right)-\frac{1}{2}\tilde{F}_{12}\left(\Box\right)-\frac{1}{2}\tilde{F}_{13}\left(\Box\right)-\frac{1}{2}\tilde{F}_{16}\left(\Box\right)+\tilde{F}_{17}\left(\Box\right)-\frac{1}{2}\tilde{F}_{20}\left(\Box\right)-\frac{1}{2}\tilde{F}_{21}\left(\Box\right)-\frac{1}{2}\tilde{F}_{22}\left(\Box\right)
\nonumber
\\
&-&\frac{1}{2}\tilde{F}_{23}\left(\Box\right)+\tilde{F}_{33}\left(\Box\right)+\frac{1}{2}\tilde{F}_{42}\left(\Box\right)-\frac{1}{2}\tilde{F}_{43}\left(\Box\right)-\frac{1}{2}\tilde{F}_{44}\left(\Box\right)-\frac{1}{2}\tilde{F}_{45}\left(\Box\right)-\frac{1}{2}\tilde{F}_{46}\left(\Box\right).
\nonumber
\end{eqnarray}
}
Explicit expressions for the purely metric functions $a,b,c,d,f$ are given in 
~\cite{Biswas:2011ar}.
\end{widetext}

%
%

\begin{thebibliography}{99}

\bibitem{Kibble:1961ba}
  T.~W.~B.~Kibble,
  J.\ Math.\ Phys.\  {\bf 2} (1961) 212.

\bibitem{Shapiro}
I. L. Shapiro, Physical aspects of the spacetime torsion,
Physics Reports, 357(2), 113-213 (2002); 
(M. Blagojevic and F. W. Hehl (eds.)). Gauge Theories of Gravitation. A Reader with Commentaries (Imperial College Press, London, 2013).


\bibitem{Tomboulis:1997gg} 
  E.~T.~Tomboulis,
  hep-th/9702146.

\bibitem{Modesto}
L. Modesto Phys.Rev. D86 (2012) 044005 

\bibitem{Biswas:2005qr} 
  T.~Biswas, A.~Mazumdar and W.~Siegel,
  JCAP {\bf 0603}, 009 (2006)
  A.~A.~Tseytlin,
  Phys.\ Lett.\ B {\bf 363}, 223 (1995)
  W. Siegel, Stringy gravity at short distances, hep-th/0309093
  

\bibitem{Biswas:2011ar} 
  T.~Biswas, E.~Gerwick, T.~Koivisto and A.~Mazumdar,
  Phys.\ Rev.\ Lett.\  {\bf 108}, 031101 (2012)



\bibitem{Koivisto1}
T.~Koivisto and G.~Tsimperis,
  arXiv:1810.11847 [gr-qc].

\bibitem{Christensen}
S.M. Christensen, J. Phys. A 13 (1980) 3001.


\bibitem{Koivisto2}
A.~Conroy and T.~Koivisto,
  EPJC {\bf 78} (2018) no.11,  923

\bibitem{constraints}
C. L\"ammerzahl,  Phys. Lett. A, 228(4-5), 223 (1997).


\bibitem{Biswas:2016etb} 
  T.~Biswas, A.~S.~Koshelev and A.~Mazumdar,
  Fundam.\ Theor.\ Phys.\  {\bf 183}, 97 (2016)
   T.~Biswas, A.~S.~Koshelev and A.~Mazumdar,
  Phys.\ Rev.\ D {\bf 95}, no. 4, 043533 (2017)


\bibitem{Ferraris}
M. Ferraris, M. Francaviglia, and C. Reina, 
Gen. Relativ. Gravit. 14, 243 (1982).

\bibitem{Biswas:2013cha} 
  T.~Biswas, A.~Conroy, A.~S.~Koshelev and A.~Mazumdar,
  Class.\ Quant.\ Grav.\  {\bf 31}, 015022 (2014)
  Erratum: [Class.\ Quant.\ Grav.\  {\bf 31}, 159501 (2014)]

\bibitem{VanNieuwenhuizen:1973fi}
  P.~Van Nieuwenhuizen,
  Nucl.\ Phys.\ B {\bf 60} (1973) 478.
T.~Biswas, T.~Koivisto and A.~Mazumdar,
  ``Nonlocal theories of gravity: the flat space propagator,''
  arXiv:1302.0532 [gr-qc].
   L.~Buoninfante,
  ``Ghost and singularity free theories of gravity,''
  arXiv:1610.08744 [gr-qc].

\bibitem{Edholm:2016hbt} 
  J.~Edholm, A.~S.~Koshelev and A.~Mazumdar,
  Phys.\ Rev.\ D {\bf 94}, no. 10, 104033 (2016)

\bibitem{Buoninfante:2018xiw} 
  L.~Buoninfante, A.~S.~Koshelev, G.~Lambiase and A.~Mazumdar,
  JCAP {\bf 1809}, no. 09, 034 (2018)
  L.~Buoninfante, et.al,
  JCAP {\bf 1806}, no. 06, 014 (2018)


\bibitem{Frolov:2015bia} 
  V.~P.~Frolov, A.~Zelnikov and T.~de Paula Netto,
  JHEP {\bf 1506}, 107 (2015)
  V.~P.~Frolov,
  Phys.\ Rev.\ Lett.\  {\bf 115}, no. 5, 051102 (2015)
  V.~P.~Frolov and A.~Zelnikov,
  Phys.\ Rev.\ D {\bf 93}, no. 6, 064048 (2016)

\bibitem{Mazumdar:2018xjz} 
  A.~Mazumdar and G.~Stettinger,
  ``New massless and massive infinite derivative gravity in three dimensions around Minkowski and in (A)dS,''
  arXiv:1811.00885 [hep-th].
  
  
\bibitem{Boos:2018bxf} 
  J.~Boos, V.~P.~Frolov and A.~Zelnikov,
  Phys.\ Rev.\ D {\bf 97}, no. 8, 084021 (2018)
  
  \bibitem{Koshelev:2018hpt}
  A.~S.~Koshelev, J.~Marto and A.~Mazumdar,
  Phys.\ Rev.\ D {\bf 98} (2018) no.6,  064023
  
  \bibitem{Koshelev:2017bxd} 
  A.~S.~Koshelev and A.~Mazumdar,
  Phys.\ Rev.\ D {\bf 96}, no. 8, 084069 (2017)

\bibitem{delaCruz-Dombriz:2018vzn}
 A.~de la Cruz-Dombriz and F.~J.~M.~Torralba,
  ``Birkhoff's theorem for general stable torsion theories,''
  arXiv:1811.11021 [gr-qc].
  
\bibitem{Buoninfante:2018xif} 
  L.~Buoninfante, et.al,
  Phys.\ Rev.\ D {\bf 98}, no. 8, 084041 (2018)





\end{thebibliography}
\end{document}